\documentclass[pra,reprint,superscriptaddress,bibnotes,floatfix]{revtex4-1}  

\usepackage[T1]{fontenc}
\usepackage[utf8]{inputenc} 

\usepackage{amsmath,amssymb,graphicx,bm,microtype} 
\usepackage[dvipsnames]{xcolor} 
\usepackage{siunitx}
\usepackage{braket} 
\usepackage[colorlinks,allcolors=blue!50!black]{hyperref} 
\usepackage[all]{hypcap}
\usepackage{cleveref}

\renewcommand{\vec}[1]{\bm{\mathrm{#1}}}
\newcommand{\ketbra}[2]{\ket{#1}\!\bra{#2}}
\newcommand{\inprod}[2]{\langle{#1}|{#2}\rangle}

\let\Re\relax \DeclareMathOperator{\Re}{Re}
\let\Im\relax \DeclareMathOperator{\Im}{Im}

\begin{document}

\title{Coherent and Controllable Enhancement of Light-harvesting Efficiency}

\author{Stefano Tomasi}
\affiliation{School of Chemistry and University of Sydney Nano Institute, University of Sydney NSW 2006, Australia}

\author{Sima Baghbanzadeh}
\affiliation{School of Physics, Institute for Research in Fundamental Sciences (IPM), P.O. Box 19395-5531, Tehran, Iran}
\affiliation{\mbox{Center for Quantum Information and Control, University of New Mexico, Albuquerque NM 87131, USA}}

\author{Saleh Rahimi-Keshari}
\affiliation{School of Physics, Institute for Research in Fundamental Sciences (IPM), P.O. Box 19395-5531, Tehran, Iran}
\affiliation{\mbox{Center for Quantum Information and Control, University of New Mexico, Albuquerque NM 87131, USA}}
\affiliation{Institut f\"{u}r Theoretische Physik, Leibniz Universit\"{a}t Hannover, 30167 Hannover, Germany}

\author{Ivan Kassal}
\email[Email: ]{ivan.kassal@sydney.edu.au}
\affiliation{School of Chemistry and University of Sydney Nano Institute, University of Sydney NSW 2006, Australia}

\begin{abstract}
Spectroscopic experiments have identified long-lived coherences in several light-harvesting systems, suggesting that coherent effects may be relevant to their performance. However, there is limited experimental evidence of coherence enhancing light-harvesting efficiency, largely due to the difficulty of turning coherences on and off to create an experimental control. Here, we show that coherence can indeed enhance light harvesting, and that this effect can be controlled. We construct a model system in which initial coherence can be controlled using the incident light, and which is significantly more efficient under coherent, rather than incoherent, excitation. Our proposal would allow for the first unambiguous demonstration of light harvesting enhanced by intermolecular coherence, as well as demonstrate the potential for coherent control of excitonic energy transfer.

\end{abstract}

\maketitle

Observations of long-lasting coherences in light-harvesting systems---such as photosynthetic pigment-protein complexes~\cite{engel2007,Panitchayangkoon2010,Collini2010,Liu2007,Hayes2013}---that were previously thought to be too noisy to support coherent effects, have raised the question of whether coherence can play a role in molecular light-harvesting  processes~\cite{engel2007,Scholes2010,Scholes2011,Kassal2013,Fassioli2013,Scholes2017,Romero2017,Shatokhin2018}. 
The question remains open, despite arguments that the observed coherences are dominantly vibrational or vibronic~\cite{Christensson2012,Tiwari2013,Thyrhaug2018,Duan2017} and that, in either event, they could not occur in nature because they could not be induced by incoherent sunlight~\cite{Jiang1991,Mancal2010,Brumer2012,Kassal2013,Dodin2016a,Brumer2018}.
Nevertheless, theoretical studies have proposed that, even in those circumstances, there are mechanisms by which coherences could enhance light-harvesting efficiencies~\cite{Dorfman2013,Svidzinsky2011,Scully2010,Scully2011,Creatore2013,Kassal2013,Leon-Montiel2014,Dodin2016,Dodin2016a,Baghbanzadeh2016,Baghbanzadeh2016_2,Oviedo-Casado2016,Fruchtman2016,Higgins2017,Tscherbul2018,Brumer2018,Brown2019,Rouse2019}.
In most of those works, as here, efficiency is defined as the probability of an excitation being successfully transferred to a target acceptor, which, in many cases, eventually leads to charge transfer or another means of harvesting the excitation energy. 
Aside from providing insight into fundamental questions on the influence of coherence in light-harvesting processes, research on this topic is also motivated by the potential application of these concepts to the design of novel artificial light-harvesting devices~\cite{Scholes2010,Romero2017}. 

Direct experimental evidence of an efficiency enhancement due to intermolecular coherence is lacking for two reasons. First, experiments so far have focused on observing coherences in isolated systems, with no acceptor for the excitations to be transferred to, making them unable to relate coherence to efficiency. Second, to ensure that a particular enhancement is due to coherence and not a confounding factor, it would be necessary to be able to switch coherence on and off without affecting other experimental variables. This kind of control is often not possible in existing light harvesting systems; for example, altering their molecular structures often causes significant changes to their overall energy landscape~\cite{Baghbanzadeh2016}. 

The only demonstrations of coherent enhancements have been experiments showing that the efficiency of excitation transfer from one molecule can be increased through adaptive optimal control~\cite{Herek:2002wy,Savolainen:2008gh}. These experiments targeted intramolecular (often vibrational~\cite{Savolainen:2008gh}) coherences within the donor, leaving the effect of intermolecular coherences on efficiency unobserved. Theoretical work has shown that multi-chromophoric light harvesting could also be controlled~\cite{Hoyer:2014bf,Caruso:2012hv}, but the final pulse sequences produced by sophisticated optimisation algorithms can be difficult to understand intuitively.

\begin{figure}[t]
\centering
\includegraphics[width=\linewidth]{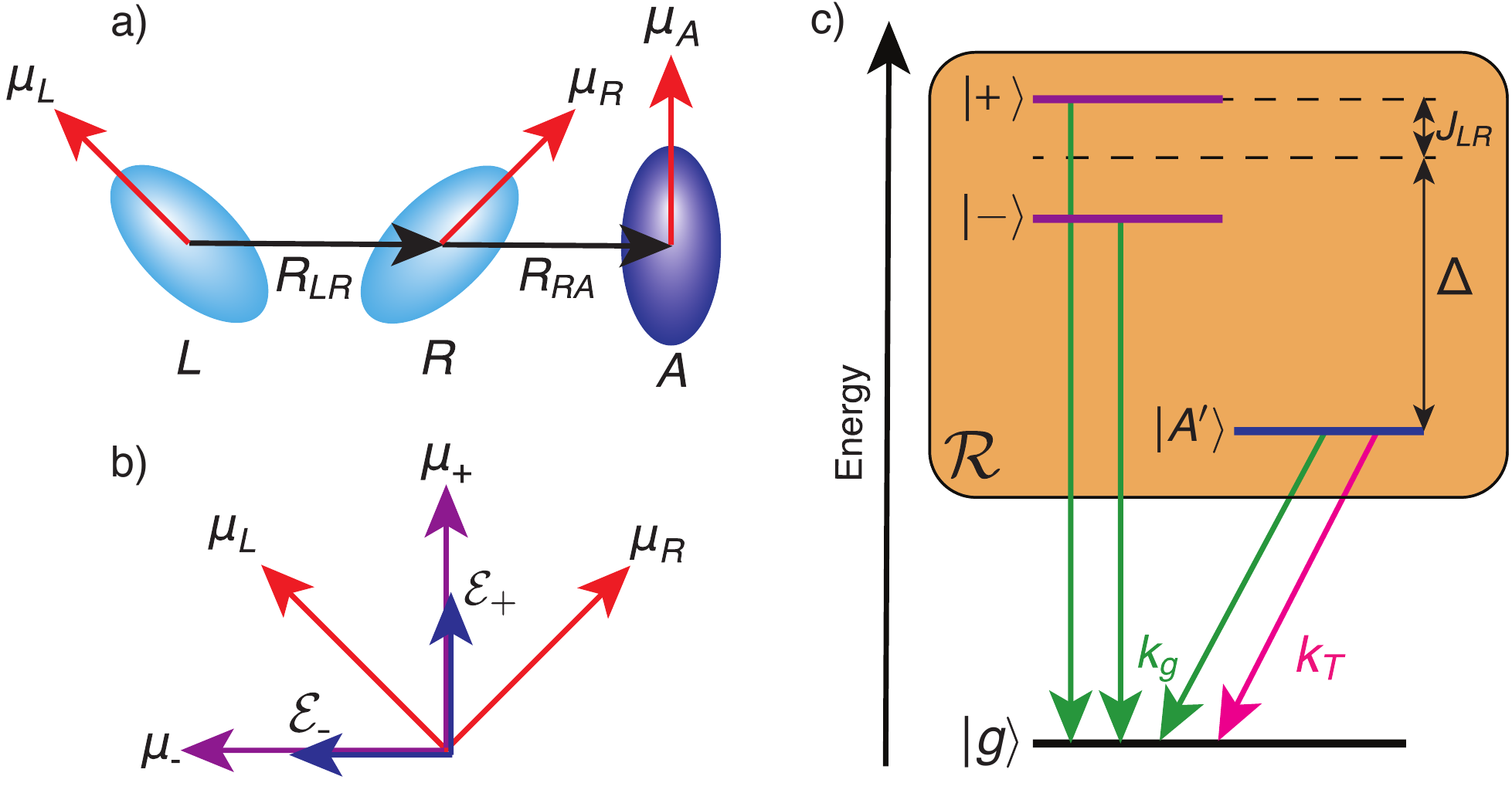}
\caption{Light-harvesting system under investigation. 
\textbf{(a)} Two identical donors ($L$ and $R$) are coupled to each other and to an acceptor ($A$). The site transition dipole moments $\mu$ lie in the plane of the paper.
\textbf{(b)} The transition dipole moments of $L$ and $R$ are compared to those of the eigenstates $\ket{+}$ and $\ket{-}$. The light is coming into the page, with light component $\mathbf{\mathcal{E}}_\pm$ assumed parallel to $\vec{\mu}_\pm$.
\textbf{(c)} The eigenstates of the model and the dissipation pathways. Dissipation within the excited-state manifold (orange) is described by a non-secular Redfield tensor $\mathcal{R}$, meaning it cannot be represented using simple rate processes. The acceptor's excited state is lower in energy compared to the donors to make the donor-to-acceptor excitation transfer energetically favourable.
} 
\label{fig:model}
\end{figure}

Here, we address the problem from the bottom up.
Instead of describing existing light-harvesting systems, our goal is to design a minimally complex light-harvesting system whose efficiency can be directly monitored and whose coherence can be externally controlled.
The ability to compare light-harvesting efficiencies in the presence and absence of coherence would permit the first definitive demonstration of light-harvesting enhanced by intermolecular coherence.

\begin{figure}[t]
\centering
\includegraphics[width=1.0\linewidth]{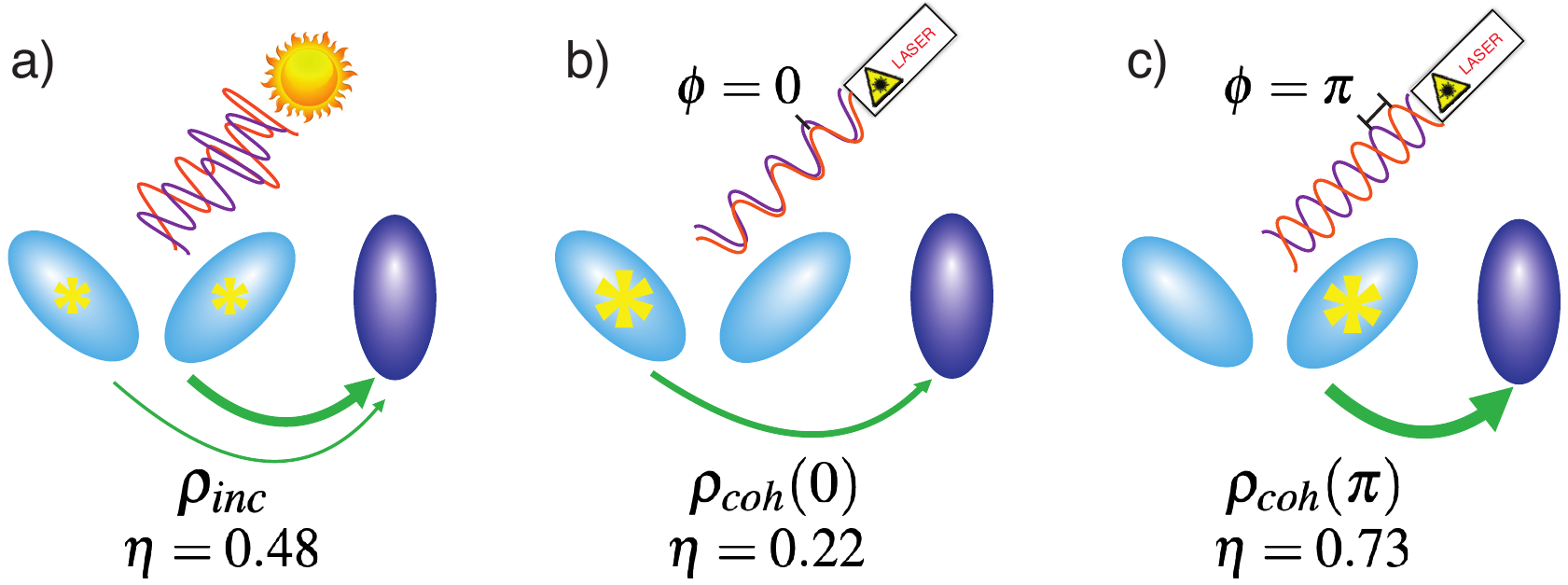}
\caption{Mechanism of coherent efficiency enhancement. 
\textbf{(a)} Incoherent light excites a mixture of donor eigenstates, while coherent light \textbf{(b--c)} can be used to address the individual $\ket{+}$ and $\ket{-}$ eigenstates and localise excitations to either donor site. The localisation is achieved by controlling only the relative phase between the two light modes and not the spectrum of the light. The donor-acceptor transfer rate---and therefore the light-harvesting efficiency $\eta$---can be significantly enhanced or diminished compared to the incoherent case if excitations are localised on $R$ or $L$ respectively. The values of $\eta$ shown were calculated using parameters described in the text.
}
\label{fig:left_right}
\end{figure}

Our system consists of two identical donor sites (e.g., molecules) and an acceptor site (Fig. \ref{fig:model}a). The acceptor's excited state is significantly red-shifted compared to the donors', ensuring that the donor-to-acceptor excitation transfer is both irreversible and spectrally resolvable. 
Excitonic coupling between the donors forms two eigenstates that are delocalised across the donor dimer and addressable using different light modes.
Using optical phase control---i.e., changing only the phases but not the intensity of the light---the system can be prepared in a wide range of coherent and incoherent initial states~\cite{shapiro_brumer,Leon-Montiel2014,Bruggemann2004,Caruso:2012hv}.
By measuring the proportion of excitations successfully transferred to the acceptor (as opposed to lost to recombination), we can compare energy-transfer efficiencies for different initial states and unambiguously demonstrate the influence of excitonic  coherence on light-harvesting efficiency.
In particular, certain superpositions of eigenstates represent excitations that are mostly localised on particular sites, allowing for efficiency enhancements if excitations are localised close to the acceptor (Fig.~\ref{fig:left_right}) ~\cite{Leon-Montiel2014,Tscherbul2018}.

\section{General model: Time evolution and efficiency}

We treat the system with a Frenkel-type (tight-binding) Hamiltonian,
\begin{equation}
H_S= \sum_u \epsilon_u \ketbra{u}{u} + \sum_{u\neq v}J_{uv}\ketbra{u}{v},
\end{equation}
where $\ket{u}$ represents an excitation localised on site $u$ with energy $\epsilon_u$ and $J_{uv}$ is the coupling between sites $u$ and $v$. We assume dipole-dipole inter-site couplings 
\begin{equation}
J_{uv}= \left[\bm{\mu}_{ug}\cdot\bm{\mu}_{vg}-3(\bm{\mu}_{ug}\cdot\bm{\hat{R}}_{uv})(\bm{\mu}_{vg}\cdot\bm{\hat{R}}_{uv})\right]/4\pi\varepsilon R_{uv}^3,
\end{equation}
where $\varepsilon$ is the dielectric constant, $\bm{\mu}_{ug}$ is the ground-to-excited-state transition dipole moment of site $u$, $\bm{R}_{uv}$ is the  distance between two sites, $R_{uv}\equiv|\bm{R}_{uv}|$ and $\bm{\hat{R}}_{uv}= \bm{R}_{uv}/R_{uv}$.

When the system is coupled to an environment, the time evolution of its reduced density operator (RDO) $\rho$ is given by the master equation \begin{equation}
\dot{\rho}= -\frac{i}{\hbar}\left[H_S,\rho\right] + \mathcal{R}\rho + \mathcal{L}_g\rho + \mathcal{L}_T\rho,
\label{eq:master}
\end{equation}
which involves three dissipators: $\mathcal{R}$ describes dissipation within the single-exciton manifold and $\mathcal{L}_g$ and $\mathcal{L}_T$ describe the relaxation of the excited states to a ground state $\ket{g}$.

Single-exciton-manifold dissipation $\mathcal{R}$ is described by coupling the excited states of the system to local nuclear degrees of freedom. The environment consists of independent baths of harmonic oscillators on each site, $H_{B}=\hbar\sum_{u,\xi}\omega_\xi b^{(u)\dag}_\xi b^{(u)}_\xi$
with $b^{(u)\dag}_\xi$ and $b^{(u)}_\xi$ the creation and annihilation operators for mode $\xi$ on site $u$.
The system-bath coupling is assumed to be linear,
$H_{SB}=\hbar\sum_{u,\xi}\omega_\xi g_\xi \ketbra{u}{u}\left(b^{(u)\dag}_\xi+b^{(u)}_\xi\right)$.
This model assumes the effect of intra-molecular vibrations dominates over inter-molecular vibrations, which is usually the case for disordered molecular aggregates~\cite{maykuhn}.

We assume that the bath affects the excited-state dynamics weakly and $\mathcal{R}$ can therefore be described using Redfield theory~\cite{maykuhn,breuer},
\begin{multline}
\mathcal{R}\rho = \sum\limits_{ab,cd} \Gamma_{ab,cd}(\omega_{dc})\left(M_{cd}\rho M_{ab} - M_{ab}M_{cd}\rho\right) \\ + \Gamma_{dc,ba}(\omega_{ab})\left(M_{cd}\rho M_{ab} - \rho M_{ab}M_{cd}\right),\label{eq:R_expanded}
\end{multline}
\begin{equation}
\Gamma_{ab,cd}(\omega) = \frac{1}{2}\sum_{u} \inprod{a}{u}\inprod{u}{b}\inprod{c}{u}\inprod{u}{d} \tilde{C}(\omega),\label{eq:Gamma}
\end{equation}
where $\ket{a},\ket{b},\ket{c}$ and $\ket{d}$ are eigenstates of $H_S$, $M_{ab}=\ketbra{a}{b}$, $\omega_{ab}$ is the transition frequency between two eigenstates $\ket{a}$ and $\ket{b}$, and where we ignored the Lamb shift caused by $H_{SB}$, which could be incorporated into $H_S$~\cite{maykuhn,breuer}.
The Fourier transform of the bath correlation function is 
$\tilde{C}(\omega)= 2\pi\omega^2(n(\omega)+1)(J(\omega)-J(-\omega))$~\cite{maykuhn},
where $n(\omega)$ is the Bose-Einstein distribution at $T=\SI{300}{K}$ and $J(\omega)= \sum_\xi g(\omega_\xi)^2 \delta(\omega-\omega_\xi)$ is the bath spectral density. Eq.~\ref{eq:Gamma} assumes identical and uncorrelated spectral densities on all sites.
Because $\tilde{C}(0)=0$ the interaction of a system with a bath of harmonic oscillators does not cause pure dephasing of system eigenstates~\cite{maykuhn}.

The remaining dissipators in Eq.~\ref{eq:master}, $\mathcal{L}_g$ and $\mathcal{L}_T$, describe exciton recombination to the ground state. $\mathcal{L}_g$ collectively describes all exciton recombination processes (whether radiative or non-radiative) where the energy is lost to the environment. We describe it using the superoperator
\begin{equation}
\mathcal{L}_g\rho= \sum_a k_g^{(a)} \left(\ketbra{g}{a}\rho\ketbra{a}{g}-\frac{1}{2}\left\{\ketbra{a}{a},\rho\right\}\right),
\end{equation}
where $\{ \cdot,\cdot \}$ denotes an anticommutator and $k_g^{(a)}$ is the recombination rate from state $a$. For the sake of simplicity, we assume that all $k_g^{(a)}$ are equal to a constant rate $k_g$. We also define a \emph{target} process that causes a particular eigenstate $\ket{T}$ to decay to $\ket{g}$ at rate $k_T$ in a way that its energy is somehow captured. This process can, for example, represent the separation of an exciton into charge carriers, or simply an external observation of a donor-acceptor transfer event.
We model the target process as
\begin{equation}
\mathcal{L}_T\rho= k_T\left(\ketbra{g}{T}\rho\ketbra{T}{g} -\frac{1}{2}\left\lbrace\ketbra{T}{T},\rho\right\rbrace\right).
\end{equation}

Distinguishing between the desirable target process and the wasteful recombination leads to a definition of efficiency as the probability of an excitation moving from $\ket{T}$ to $\ket{g}$ during the lifetime of the excitation, given by
\begin{equation}
\eta= \int^\infty_0 k_T \bra{T}\rho(\tau)\ket{T}d\tau.
\end{equation}

Because the efficiency depends only on populations, the only way for coherences to influence the efficiency is if they can affect the populations. Therefore, the terms that cause coherence-to-population transfer in eq.~\ref{eq:R_expanded} are important. Each term in eq.~\ref{eq:R_expanded} leads to time evolution that oscillates at a frequency $\omega_{ab}-\omega_{dc}$, where $\omega_{ab}= (E_a-E_b)/\hbar$. Often, these oscillations are sufficiently fast for the influence of particular terms to average to zero on time scales of interest. This motivates the widely used secular approximation, where all terms with $|\omega_{ab}-\omega_{cd}|\neq 0$ are discarded~\cite{maykuhn,breuer}, eliminating all transfers between populations and coherences and decoupling population and coherence dynamics.
However, if two levels are nearly degenerate, some terms connecting them may oscillate slowly enough to significantly affect population dynamics, making it unsafe to discard them~\cite{Dodin2017,Tscherbul2014}. In our case, with an efficiency depending on populations, non-secular effects, found in the limit of nearly degenerate states, are essential for coherent efficiency enhancements.

\begin{figure*}[tb]
\centering
\includegraphics[width=\textwidth]{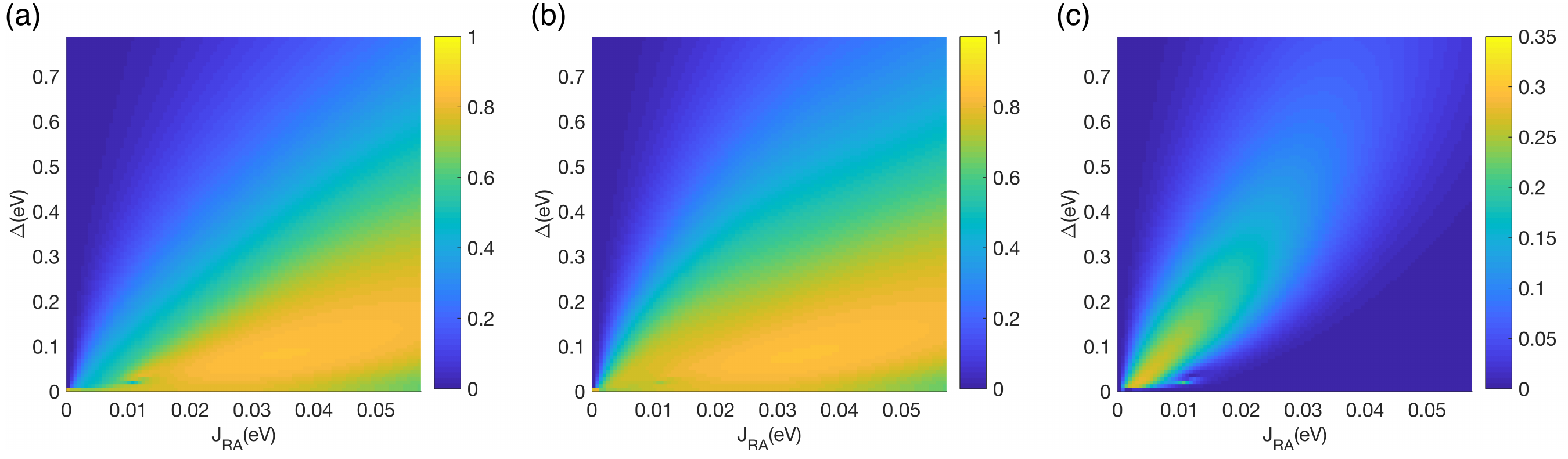}
\caption{Light-harvesting efficiency for \textbf{(a)} excitations caused by incoherent light, \textbf{(b)} coherent excitations with phase $\phi=\pi$ and \textbf{(c)} the difference between the two, for a range of donor-acceptor detunings $\Delta$ and couplings $J_{RA}$ between $\ket{R}$ and the acceptor, evaluated with the full, non-secular Redfield tensor $\mathcal{R}$. The coupling between the two donors is fixed at $J_{LR}=\SI{1.4}{meV}$. The coherent enhancement of populations of $\ket{R}$ enhances the efficiency the most when neither $\Delta$ nor $J_{RA}$ is too large.}
\label{fig:comp_surf}
\end{figure*}

\section{Three-site minimal model: Eigenstates and state preparation} 

To give the simplest example of controllable efficiency enhancement, we consider a system of three sites, the left donor $\ket{L}$, the right donor $\ket{R}$, and the acceptor $\ket{A}$. We assume the two donors have degenerate excited states, so $\epsilon_L=\epsilon_R\equiv\epsilon_D$ and we let $\epsilon_A= \epsilon_D-\Delta$, where $\Delta$ is an energy detuning. For $\Delta\gg J_{LR}$, by diagonalising $H_S$ we obtain two eigenstates that are approximately delocalised exclusively across the two donors,
\begin{subequations}\label{eq:donor_eig}
\begin{align}
\ket{+}&\approx \sqrt{p_1}\ket{L}+\sqrt{p_2}\ket{R} \\
\ket{-}&\approx \sqrt{p_2}\ket{L}-\sqrt{p_1}\ket{R}
\end{align}
\end{subequations}
with energies $E_{\pm}\approx \epsilon_D\pm J_{LR}$ and $p_1+p_2=1$.
In general, these two eigenstates also overlap with $\ket{A}$; however, when $\Delta\gg J_{LR}$, this overlap is small enough that we can assume the eigenstates are contained within the donor dimer, and we refer to them as donor eigenstates. In addition, the third energy eigenstate $\ket{A'}$ coincides with $\ket{A}$, up to a small perturbation, and we refer to it as the acceptor eigenstate. The eigenstates are shown in Fig.~\ref{fig:model}c.
Nevertheless, to ensure accurate rates in the Redfield tensor (eq.~\ref{eq:R_expanded}), the calculations below include the small overlaps of the donor eigenstates with $\ket{A}$ and of $\ket{A'}$ with the donor sites.

To show that coherences between eigenstates can affect the efficiency, we consider cases where $\omega_{+-}$ is smaller than the donor-to-acceptor transfer rate.
In this regime, non-secular terms oscillating at frequencies less than $2|\omega_{+-}|$ can have a significant effect on system dynamics.

To control the optically prepared initial state, the individual eigenstates should be addressable by different optical modes. In principle, the modes could be different frequencies, but the significant broadening of our eigenstates by their fast decay may make it impossible to resolve them spectrally. This difficulty can be overcome by also considering optical modes with different polarisation.

The initial state depends on whether the exciting light is coherent or incoherent (or, for polarisation, polarised or unpolarised)~\cite{Jiang1991,Mancal2010,Brumer2012}. System-light coupling is weak in molecular light-harvesting systems and, therefore, treatable a first-order perturbation~\cite{Jiang1991,Mancal2010,Brumer2012}.
Consequently, the total excited-state population is always much smaller than the ground-state population. However, in the following, we are only interested in the excited-state populations after an initial light pulse. Therefore, we exclude initial ground-state populations and normalise initial excited states so their populations add to unity.

Weak polarised coherent light prepares the excited state~\cite{breuer,Bruggemann2004,shapiro_brumer,Jiang1991,Leon-Montiel2014}
\begin{equation}
\ket{\psi_{\text{coh}}}= \frac{1}{\sqrt{\mathcal{N}}}\sum\limits_a |\vec{\mu}_{ag}\cdot\vec{\mathcal{E}}_a|e^{i\phi_a}\ket{a},
\label{eq:coh}
\end{equation}
where $\vec{\mathcal{E}}_a$ and $\phi_a$ are the electric field amplitude and the phase of the light mode exciting eigenstate $\ket{a}$, $\vec{\mu}_{ag}$ is the transition dipole moment for the $g\to a$ transition and $\mathcal{N}$ is a normalisation factor.
The transition dipole moments of the eigenstates are linear combinations of the site-basis transition dipoles,
$\bm{\mu}_{ag} = \sum_u \inprod{u}{a} \bm{\mu}_{ug}$.
In order to individually address the $\ket{+}$ and $\ket{-}$ states, we choose donor sites with perpendicular dipole moments of equal magnitude. This arrangement results in donor eigenstates whose dipole moments are also perpendicular and of equal magnitude, making them addressable using separate polarisation modes of the light (Fig.~\ref{fig:model}b).

By contrast, incoherent light excites a mixed state~\cite{Jiang1991,Mancal2010,Brumer2012}. If the eigenstates are spectrally distinguishable, the state will be a mixture of eigenstates; by contrast, coherences between eigenstates can occur when two eigenstates overlap spectrally (due to a finite linewidth), allowing them to couple to the same light mode(s)~\cite{Dodin2016,Dodin2016a,Scully2011,Scully2010,Svidzinsky2011,Dorfman2013,Tscherbul2014}.
While the donor eigenstates in our system are nearly degenerate, and therefore not perfectly spectrally distinguishable, the orthogonality of their dipole moments implies that the two donor eigenstates couple to light modes with different polarisations~\cite{Tscherbul2014}. In unpolarised light, these modes act as uncorrelated light sources and do not induce coherences.

Because fully incoherent light is stationary, it, strictly speaking, exists only as a continuous-wave process~\cite{Chenu2016,Brumer2018} (various coherent artefacts are induced if incoherent light is suddenly switched on~\cite{Tscherbul2015,Tscherbul2014,Grinev2015,Dodin2016,Dodin2016a,Olsina2014,Brumer2018}). However, the efficiency of light harvesting in continuous-wave incoherent light is equal to the efficiency given a particular transient initial state \cite{Jesenko2013}. In our case, the equivalent initial state is
\begin{equation}
\rho_{\text{inc}}= \frac{1}{\mathcal{N}}\sum_a |\mu_{ag}|^2\overline{\mathcal{E}^2_a}\ket{a}\!\bra{a},
\end{equation}
where $\overline{\mathcal{E}^2_a}$ is the ensemble mean-square electric field intensity of the mode $\mathcal{E}_a$.

To maximise coherences under coherent excitations, we consider light sources which excite equal populations in $\ket{+}$ and $\ket{-}$.
We also assume the target process occurs via the acceptor eigenstate $\ket{A'}$ (i.e., $\ket{T}=\ket{A'}$).
Finally, we assume no direct excitation of $\ket{A'}$, which could trivially contribute to the efficiency. Practically, this would correspond to an optical excitation with no field component resonant with this state. This ensures that the target process efficiency measures successful transfers from donor states to the acceptor state.

The initial states of the system are then 
\begin{equation}
\rho_{\text{inc}}= \frac{1}{2}\left(\ketbra{+}{+}+\ketbra{-}{-}\right), \label{eq:incoh_dens}
\end{equation}
for incoherent (unpolarised) excitation, and
\begin{subequations}
\label{eq:coh_excit}
\begin{align}
\rho_{\text{coh}}(\phi)&= \ketbra{\psi_{\text{coh}}(\phi)}{\psi_{\text{coh}}(\phi)}, \\
\ket{\psi_{\text{coh}}(\phi)}&= \frac{1}{\sqrt{2}}\left(\ket{+}+e^{i\phi}\ket{-}\right),\label{eq:coh_wavefunc}
\end{align}
\end{subequations}
for coherent (polarised) excitation with relative phase $\phi=\phi_- - \phi_+$ between the two light modes. 

In the limit of eq.~\ref{eq:donor_eig}, populations on sites $L$ and $R$, given initial state $\ket{\psi_\text{coh}(\phi)}$, are
\begin{subequations}\label{eq:local}
\label{eq:localise}
\begin{align}
\rho_{LL}&\approx \frac{1}{2}+\sqrt{p_1p_2}\cos\phi, \label{eq:local_L} \\
\rho_{RR}&\approx \frac{1}{2}-\sqrt{p_1p_2}\cos\phi. \label{eq:local_R}
\end{align}
\end{subequations}
Initial excitations can be significantly localised on $L$ or $R$ through choice of $\phi$, especially for small $J_{RA}$, when $p_1\approx p_2\approx \frac12$. In particular, population on $R$ is maximised at $\phi=\pi$.

\section{Results and discussion}
For concreteness, we consider donors with energies $\epsilon_D = \SI{2.1}{eV}$ and all three sites with transition dipole moments $\mu_{ug} = \SI{7}{D}$, with geometry shown in Fig.~\ref{fig:model}.
The separation between the two donors was fixed at $R_{LR}=\SI{2}{nm}$, corresponding to $J_{LR} = \SI{1.4}{meV}$ and ensuring that $|\omega_{+-}|$ is significantly smaller than donor-acceptor transfer rates.
The bath was assumed to have a Debye spectral density~\cite{maykuhn,Pachon2017},
$\omega^2J(\omega)= \theta(\omega)2\Lambda\omega_D\omega/\hbar(\omega^2+\omega_D^2),$
with reorganisation energy $\Lambda = \SI{140}{cm^{-1}}$ and Debye frequency $\omega_D = \SI{100}{cm^{-1}}$, where $\theta(\omega)$ is the Heaviside step function.
We used target rate $k_T = \SI{300}{ns^{-1}}$ and recombination rate $k_g = \SI{50}{ns^{-1}}$.

We used eq.~\ref{eq:master}, with the full, non-secular Redfield tensor of eq.~\ref{eq:R_expanded}, to evolve three initial states: the coherent states $\rho_\text{coh}(0)$ and $\rho_\text{coh}(\pi)$, and the incoherent state $\rho_\text{inc}$. Importantly, eigenstate populations are initially identical across the three cases, and all differences in dynamics and efficiencies are caused by the coherences.

The efficiencies in Fig.~\ref{fig:left_right} show that initial coherence can profoundly affect the efficiency. In particular, coherently increasing populations on $R$ increases the efficiency by starting the excitation closer to the acceptor. 
These efficiencies were all computed for $\Delta= 60J_{LR}$ and with $R_{RA}$ chosen so that $J_{RA}=6J_{LR}$.
The observed enhancement is an example of environment-assisted single-photon coherent phase control~\cite{Brumer:1989ii,Spanner:2010hd,Arango:2013ih} because it depends only on the phases of the light modes (the intensities are the same in all three cases) and because it relies on the relaxation of donor states to the acceptor.
In this example, the difference in efficiency between $\ket{\psi_\text{coh}(\pi)}$ and $\rho_{\text{inc}}$ is 25 percentage points. The maximum enhancement would be 50 percentage points (a doubling from 50\% to 100\%), because the efficiency of the incoherent excitation is always the average of the two coherent efficiencies. This is because the efficiency is a linear function of the initial RDO and $\rho_\text{inc}= (\rho_\text{coh}(0)+\rho_\text{coh}(\pi))/2$. 

To explore the limits of coherent efficiency enhancements, we simulated the system for a range of $\Delta$ and $R_{RA}$, while holding $R_{LR}$ fixed.
Fig.~\ref{fig:comp_surf} compares incoherent efficiencies (Fig.~\ref{fig:comp_surf}a) with those of coherent excitations with phase $\phi=\pi$ (Fig.~\ref{fig:comp_surf}b). 
For simplicity, the results are shown as functions of $J_{RA}$ instead of $R_{RA}$; in all cases, $J_{LA}$ is much less than $J_{RA}$ and has a minor effect on the efficiency.
Fig.~\ref{fig:comp_surf}c shows the distinct region where the coherent efficiency can exceed the incoherent one by as much as 30 percentage points. By contrast, when $\Delta$ is small and $J_{RA}$ large, Fig.~\ref{fig:comp_surf}a and~\ref{fig:comp_surf}b show that donor-to-acceptor transfer is fast enough for efficiencies to be large for both excitation conditions, preventing a large enhancement. On the other hand, when $\Delta$ is large and $J_{RA}$ small, donor-to-acceptor coupling is too small for transfer rates to compete with the recombination rate, giving a low efficiency regardless of initial state.

To gain a better intuition on how coherences lead to increases in efficiency, we derived a simplified master equation using a partial secular approximation, by discarding only those Redfield terms that are guaranteed to be rapidly oscillating. This model is presented in the Appendix, which also identifies the non-secular terms that most significantly affect the efficiency.

\section{Conclusion}

In summary, we have shown that excitonic coherences can significantly affect energy-transfer efficiency in a light-harvesting system. The coherences can be controlled by controlling the coherence of the exciting light; compared to incoherent excitation, engineered coherent light can double the light-harvesting efficiency for a dimeric donor. Our model could easily be generalised to larger systems, in which the enhancement could be even larger because incoherent excitations could be spread across more sites, making the effects of coherent localisation more pronounced. The particular parameter regimes we explored were chosen to be realisable in engineered nanostructures, providing a platform for the development of new, quantum-inspired light-harvesting technologies. These regimes could be treated by the weak-coupling Redfield theory, and we leave to future work the extension of our results to systems with stronger system-bath coupling.

\begin{acknowledgments}
We thank Andrew Doherty, Jacob Krich, Albert Stolow, and Joel Yuen-Zhou for valuable discussions.
S.T. and I.K. were supported by the Westpac Bicentennial Foundation through a Westpac Research Fellowship, by an Australian Government Research Training Program (RTP) Scholarship, and by a scholarship and a Grand Challenge project from the University of Sydney Nano Institute.
S.B. was supported in part by the U.S. National Science Foundation, Grant No. PHY-1630114. S.R.-K. acknowledges financial support from the German Academic Exchange Service (DAAD) and Iran's National Elites Foundation (INEF).
\end{acknowledgments}

\section*{Appendix: Partial secular approximation}

To obtain more intuition about the system dynamics and how the presence of coherence ultimately affects efficiency, we can derive a simplified master equation through a partial secular approximation.
The ordinary secular approximation eliminates all terms that transfer between populations and coherences, resulting in decoupled population and coherence dynamics.
Instead, we choose to retain all non-secular terms oscillating at frequencies less than $2|\omega_{+-}|$, as these are oscillating slowly enough for their effects to be significant.

\begin{figure*}[tb]
\centering
\includegraphics[width=0.75\textwidth]{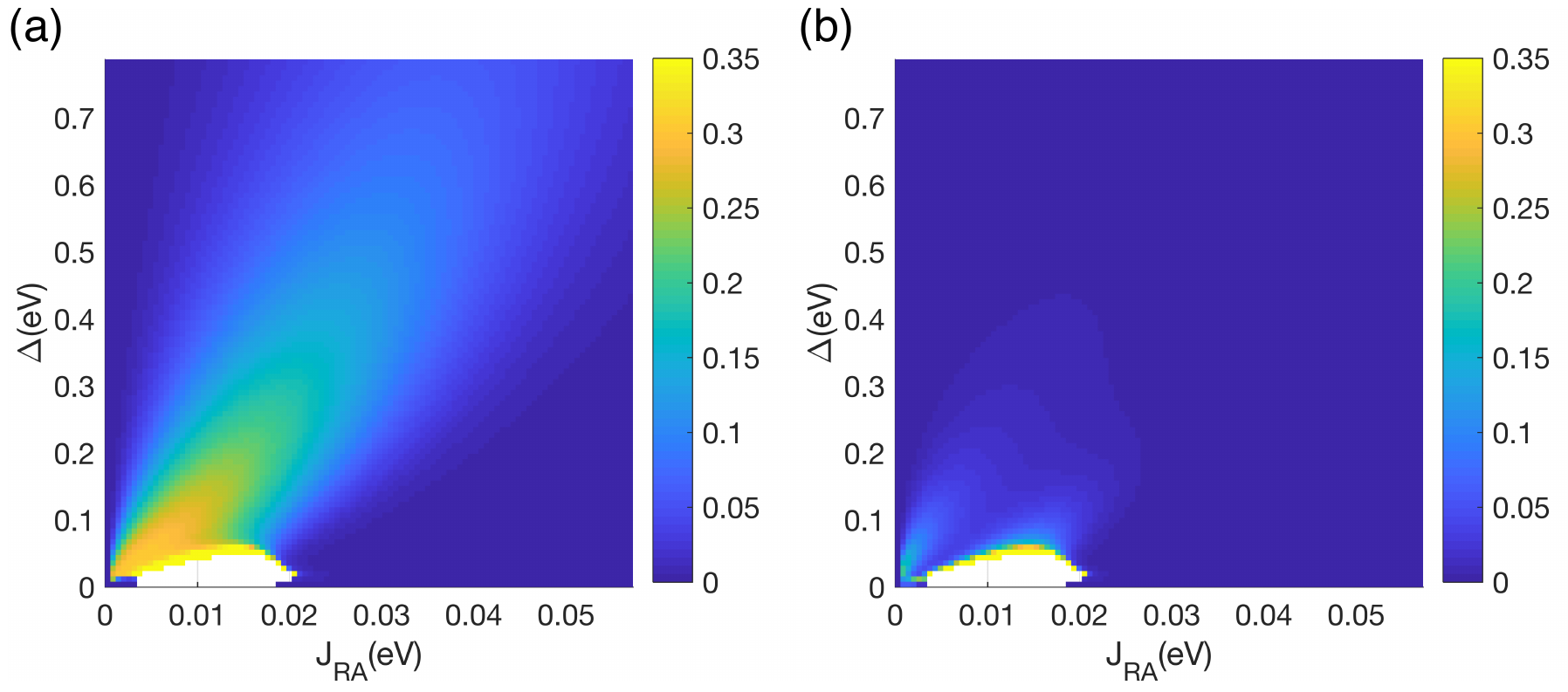}
\caption{\textbf{(a)} The behaviour in Fig.~3 is well reproduced by a simplified master equation (eq.~\ref{eq:lindblad_master}). \textbf{(b)} The difference between the two models (Fig. 3c and \ref{eq:lindblad_master}a) is small, except when $\Delta$ and $J_{RA}$ are both small, which is when the secular approximation on the acceptor state fails, sometimes causing non-physical behaviour (white region).}
\label{fig:lindblad_compare}
\end{figure*}

All other non-secular terms---namely those connecting populations to coherences involving the acceptor---oscillate quickly and can be neglected, i.e., we carry out a secular approximation on acceptor states.
After this approximation, the $\mathcal{R}$-induced evolution of each RDO element is
\begingroup
\allowdisplaybreaks
\begin{subequations}
\label{eq:lindblad_master}
\begin{align}
(\dot{\rho}_{++})^\mathcal{R}=& \sum\limits_{a=-,A'} \left(k_{+a}\rho_{aa} -k_{a+}\rho_{++}\right) -\alpha\text{Re}\left[\rho_{+-}\right],\label{eq:lindblad_plus}\\
(\dot{\rho}_{--})^\mathcal{R}=& \sum\limits_{a=+,A'} \left(k_{-a}\rho_{aa} -k_{a-}\rho_{--}\right) -\alpha\text{Re}\left[\rho_{+-}\right],\label{eq:lindblad_minus}\\
(\dot{\rho}_{A'A'})^\mathcal{R}=& \sum\limits_{a=+,-} \left(k_{A'a}\rho_{aa} -k_{aA'}\rho_{A'A'}\right) +2\alpha\text{Re}\left[\rho_{+-}\right],\label{eq:lindblad_A}\\
(\dot{\rho}_{+-})^\mathcal{R}=& -\frac{1}{2}\left(\sum\limits_{a=-,A'}k_{a+}+\sum\limits_{b=+,A'}k_{b-}\right)\rho_{+-} \nonumber \\
&+\frac{1}{2}\left(k_{+-}+k_{-+}\right)\rho_{-+} \nonumber \\
&+\frac{\alpha}{2}(2e^{-\hbar\omega_{DA'}/k_BT}\rho_{A'A'}-\rho_{++}-\rho_{--}) \nonumber \\
&+\beta\left(\rho_{++}-e^{-\hbar\omega_{+-}/k_BT}\rho_{--} \right),\label{eq:lindblad_coh}
\end{align}
\end{subequations}
\endgroup
where the population transfer rate from $\ket{a}$ to $\ket{b}$ is $k_{ba}=2\Gamma_{ab,ba}(\omega_{ab})$, $\omega_{DA'}=(\omega_{+A'}+\omega_{-A'})/2$, $\alpha=2\Gamma_{+A',A'-}(\omega_{DA'})$ and $\beta= \Gamma_{++,-+}(\omega_{+-})-\Gamma_{--,-+}(\omega_{+-})$.
We have also assumed that $\tilde{C}(\omega)$ is slowly varying over the interval $[\omega_{-A'},\omega_{+A'}]$ and can be replaced with the constant $\tilde{C}(\omega_{DA'})$. The first term in each of these equations contains the secular incoherent rates, while the remaining, non-secular terms account for coherent effects that are non-negligible in the limit of small $\omega_{+-}$.

Similar results to Fig.~3 can be obtained by propagating the approximate master equation in eq.~\ref{eq:lindblad_master}, as shown in Fig.~\ref{fig:lindblad_compare}.
Across most of the parameter space, there is little difference between the estimated efficiency enhancements obtained from the full Redfield tensor and the approximate model, validating the simpler eq.~\ref{eq:lindblad_master} as a way to understand the origin and limitations of coherent efficiency enhancements.

The cause of efficiency enhancement are population transfers from the donor states to the acceptor that are mediated by the non-secular terms (those proportional to $\alpha$) in eqs.~\ref{eq:lindblad_plus},~\ref{eq:lindblad_minus} and~\ref{eq:lindblad_A}.
In our case $\alpha< 0$, so a negative $\Re[\rho_{+-}]$ causes a decrease in donor populations and an increase in acceptor populations, while a positive $\Re[\rho_{+-}]$ has the opposite effect.
Since $\Re[\rho_{+-}]$ is negative when $\phi=\pi$, observed donor-acceptor transfer rates are fastest when populations at $R$ are maximised.
Furthermore, the sum of the additional terms is always 0, ensuring that eq.~\ref{eq:lindblad_master} is trace preserving.

In addition, because $\rho_{+-}-\rho_{-+}= 2i\Im[\rho_{+-}]$, the dephasing terms proportional to $k_{+-}$ in eq.~\ref{eq:lindblad_coh} matter only when $\rho_{+-}$ has an imaginary component.
Since our initial states have real coherences, in the limit $\omega_{+-} \ll (k_{A'+}+k_{A'-})/2 $, the coherence $\rho_{+-}$ oscillates too slowly for its imaginary component to gain significant magnitude and cause notable dephasing before the excitation transfers to the acceptor.
Therefore, and due to the absence of pure dephasing, the coherences survive and maintain a positive real part long enough for the enhancements proportional to $\alpha$ to be significant.

The approximate eq.~\ref{eq:lindblad_master} fails in the lower left part of Figure~\ref{fig:lindblad_compare}. This region is where the secular approximation with respect to the acceptor site fails. We assumed in eq.~\ref{eq:lindblad_master} that terms oscillating at frequency $|\omega_{\pm A'}|$ can be discarded due to their rapid oscillation, but this assumption fails when $\Delta$ is small. In some cases, indicated with the white region in Fig.~\ref{fig:lindblad_compare}, an efficiency could not be computed because the failure of the approximation led to populations becoming unphysical (either negative or greater than 1).


%

\end{document}